\documentclass[article]{article}
\usepackage{graphicx}
\usepackage{amsmath}
\usepackage{amssymb}
\usepackage{mathrsfs}
\usepackage[numbers,sort&compress]{natbib}
\usepackage{amsthm}
\usepackage{pgfplots}

\begin{document}

\baselineskip=20pt

\title{ The   Riemann-Hilbert approach    to   focusing Kundu-Eckhaus equation with nonzero boundary conditions   }
\author{Li-Li Wen$^1$,  En-Gui Fan$^1$\thanks{\ Corresponding author and email address: faneg@fudan.edu.cn } }
\footnotetext[1]{ \  School of Mathematical Sciences, Fudan University, Shanghai 200433, P.R. China.}

\date{}

\maketitle
\begin{abstract}
\baselineskip=20pt
In this article, we focus on investigating the focusing Kundu-Eckhaus equation  with nonzero boundary condition.
A appropriate two-sheeted Riemann surface is  introduced  to  map  the spectral parameter $k$ into a single-valued parameter $z$.
 Starting from the Lax pair of  Kundu-Eckhaus equation,
 two kind of  Jost solutions are construed.   Further their  asymptotic,  analyticity,  symmetries as well as  spectral  matrix  are detailed analyzed.
It is shown that  the solution of  Kundu-Eckhaus equation   with nonzero boundary condition
 can characterized  with a  matrix Riemann-Hilbert problem. Then  a  formula of $N$-soliton solutions is derived
  by solving Riemann-Hilbert problem.  As applications,   the first-order explicit  soliton solution is obtained.
\\
\\
\\Keywors: the focusing Kundu-Eckhaus equation; nonzero boundary conditions; Riemann-Hilbert problem; soliton solution.
\end{abstract}

\newpage

\section{Introduction}
It is well-known that  the nonlinear Schr$\ddot{o}$dinger equation   plays an important role and achieved great success in physical fields such as  nonlinear optics, nonlinear water
 waves,  plasma physics and so on.  But Schr$\ddot{o}$dinger equation with  higher-order nonlinear terms, such as the self-steepening and self-frequency shift, has an significant effect in specific physical background, for example, optic fiber communication and  Bose-Einstein condensates \cite{ky1985, kl2006}.

The modified Gross-Pitaevskii equation \cite{kvr2010}
\begin{equation}
iu_{t}+u_{xx}+2f(t)|u|^{2}u+\frac{1}{2}\delta^{2}x^{2}u+\kappa_{1}|u|^{4}u+i\kappa_{2}(|u|^{2})_{x}u=0 \label{mgp}
\end{equation}
 was firstly proposed by Wadati for describing the interactions between two-body and three-body on the condensates. The variable parameters $\kappa_{1}$ and $\kappa_{2}$ are constants. The two-body scattering lengths $f(t)$ can be adjusted by Feshbach resonance. Neither $\kappa_{1}=0$ nor $\kappa_{2}=0$, equation (\ref{mgp}) is suitable for Bose-Einstein condensates with higher densities. Reversely, if $\kappa_{1}=0$ and $\kappa_{2}=0$, equation (\ref{gp}) only for lower densities in Bose-Einstein condensates. This situation can be described by Gross-Pitaevskii equation \cite{pcj2004}
\begin{equation}
iu_{t}+u_{xx}+2f(t)|u|^{2}u+\frac{1}{2}\delta^{2}x^{2}u=0.\label{gp}
\end{equation}
 If $f(t)\in\mathbb{R}$ and independent of $t$, the equation (\ref{mgp}) reduce to  the Kundu-Eckhaus (KE) equation \cite{ka1984}
\begin{equation}
iu_{t}+u_{xx}-2\sigma|u|^{2}u+4\beta^{2}|u|^{4}u+4i\beta\sigma(|u|^{2})_{x}u=0, \label{KE1}
\end{equation}
 with $u(x,t)$ being  the complex potential function of spatial $x$ and temporal $t$ $u(x,t): \mathbb{R}^{2}\mapsto\mathbb{C}$. This equation is called  the defocusing KE equation as   $\sigma=1$
 and focusing KE equation as  $\sigma=-1$. The KE  equation (\ref{KE1}) was firstly put forward by Kundu to investigate the Landau-Lifshitz equations and derivative nonlinear Schr$\ddot{o}$dinger type equations \cite{ka1984}.
 For the special case $\beta=0$,  the  KE equation (\ref{KE1}) reduces to nonlinear Schr$\ddot{o}$dinger  equation.
 In recent years, the  KE equation (\ref{KE1})  has been investigated  via different methods, for example,
the long time asymptotic \cite{zqz2018,wds2018,wds2019,gbl2018},  the higher-order rogue wave solutions \cite{wx2014},  rogue waves in a chaotic wave field \cite{bc2016}, the Darboux transformation   \cite{qd2015,gbl2012},  integrable  discretizations  \cite{dl2009}.   Recently,    the KE  equation with zero boundary conditions was investigated  by using  Riemann-Hilbert method     \cite{wds2018}.

To our  knowledge,  there is still no work on investigating   KE equation  (\ref{KE1})  with nonzero boundary conditions by using inverse scattering transformation
or Riemannn-Hilbert method. In this paper, we would like to  investigate the soliton solution of  focusing  KE equation (\ref{KE1})   with the following nonzero boundary conditions
\begin{equation}
\lim_{x\rightarrow\pm\infty}u(x,t)=u_{\pm}e^{2it(q_{0}^{2}-2\beta^{2}q_{0}^{4})+i\theta_{\pm}},\label{condition}
\end{equation}
where $|u_{\pm}|=u_{0}>0$ and $u_{0}$ and $\theta_{\pm}$ are constants,  $\theta_{\pm}$ is the arguments of $q_{\pm}$.

As we all known, the inverse scattering transform method plays an important role for finding the exact solutions of completely integrable systems \cite{csg1967}.  The Riemann-Hilbert method as a new version of inverse scattering transform method streamline the research process and preferred by researchers. And Riemann-Hilbert method can be used to investigate the soliton solutions \cite{yjk2010} and the long-time  asymptotic of integrable systems \cite{zqz2018}. Especially, in recent years, it has become a hot topic to  investigate  integrable systems with   nonzero boundary conditions  \cite{pb2006,df2013,kdk2016,kd2015,pb2015,bp2015,cv2015,gb2015,gb2016,gb2014,pm2017,ls2018}.

This work is organized as follows.  In  section 2,  we   analyze the  spectral problem  via introduce a transformation. And we introduced a appropriate Riemann surface for the single-valued function of the spectral parameter, that is $k$-plane mapped into $z$-plane. Section 3-5, we obtained the asymptotic, analyticity and symmetries of Jost solution and scattering matrix,
which are used to get a Riemann-Hilbert problem in section 6.
 Section 7, we analyze the discrete spectrum and residue conditions which are used to solve  the Riemann-Hilbert problem.
Section 8, we establish  the connection between the solution of KE equation and the solution of Rimann-Hilbert problem.
A formula of  the $N$-soliton solution of KE equation  is obtained by using the Rimann-Hilbert problem.
 A conclusion  is given in  section 9.

\section{Spectral Analysis}

It is well-known that    the focusing KE equation ($\sigma=-1$) admits Lax pair
\begin{equation}
(\partial_{x}-\mathcal{U})\phi=0,\ \  (\partial_{t}-\mathcal{V})\phi=0,\label{laxpair}
\end{equation}
where
\begin{eqnarray}
&&\mathcal{U}=-ik\sigma_{3}+i\beta |q|^{2}\sigma_{3}+U, \hspace{0.5cm}  U=\left(
\begin{array}{cc}
   0& u \\
 -\overline{u} & 0
\end{array}\right), \ \ \ \sigma_{3}=\left(
\begin{array}{cc}
  1 & 0 \\
  0 & -1
\end{array}\right)\nonumber\\
&&\begin{split}\mathcal{V}=&-2ik^{2}\sigma_{3}+2kU+2\beta|u|^{2}U+i\left(|u|^{2}+4\beta^{2}|u|^{4}\right)\sigma_{3}-iU_{x}\sigma_{3}\\
                 &-\beta\left(UU_{x}-U_{x}U\right),
\end{split}
\nonumber\end{eqnarray}
where $k$ is the spectral parameter and $\overline{u}$ denotes the complex conjugation of $u$.

For convenience of using  Riemann-Hilbert method,  we first deal with Lax par (\ref{laxpair}) and the boundary condition (\ref{condition}).
By making  transformation
\begin{equation}
q(x,t)=u(x,t)e^{-2it(q_{0}^{2}-2\beta^{2}q_{0}^{4})},\nonumber
\end{equation}
the focusing KE equation can be reduced to the form
\begin{equation}
iq_{t}+q_{xx}+2(|q|^{2}+2\beta^{2}|q|^{4}-q_{0}^{2}-2\beta^{2}q_{0}^{4})q-4i\beta(|q|^{2})_{x}q=0, \quad \beta\in\mathbb{R},\label{KE}
\end{equation}
and corresponding boundary condition (\ref{condition}) becomes
\begin{equation}
\lim_{x\rightarrow\pm\infty}q(x,t)=q_{\pm}=q_{0}e^{i\theta_{\pm}},\nonumber
\end{equation}
where we have denote  $u_{\pm}$ and  $u_{0}$ as  $q_{\pm}$  and  $ q_0$ respectively.

The equation (\ref{KE}) admits the following Lax pair
\begin{subequations}
\begin{align}
&(\partial_{x}-\mathcal{U})\phi=0,\label{LaxU}\\
&(\partial_{t}-\mathcal{V})\phi=0,\label{LaxV}
\end{align}\label{Lax1}
\end{subequations}
where
\begin{eqnarray}
&&\mathcal{U}=-ik\sigma_{3}+i\beta |q|^{2}\sigma_{3}+Q, \hspace{0.5cm}  Q=\left(
\begin{array}{cc}
   0& q \\
 -\overline{q} & 0
\end{array}\right),\nonumber\\
&&\begin{split}\mathcal{V}=&-2ik^{2}\sigma_{3}+2kQ+2\beta|q|^{2}Q+i\left(|q|^{2}+4\beta^{2}|q|^{4}-q_{0}^{2}-2\beta^{2}q_{0}^{4}\right)\sigma_{3}\\
        &-iQ_{x}\sigma_{3}-\beta\left(QQ_{x}-Q_{x}Q\right).
\end{split}
\nonumber\end{eqnarray}

Let $x\rightarrow\pm\infty$  and by using boundary condition (\ref{condition}),  we  obtain  the  following asymptotic   Lax pair
\begin{subequations}
\begin{align}
&(\partial_{x}-\mathcal{U}_{\pm})\phi=0, \quad \mathcal{U}_{\pm}=\lim_{x\rightarrow\pm\infty}\mathcal{U}=\left(-ik+i\beta q_{0}^{2}\right)\sigma_{3}+Q_{\pm},\label{LaxU2}\\
&(\partial_{t}-\mathcal{V}_{\pm})\phi=0, \quad \mathcal{V}_{\pm}=\lim_{x\rightarrow\pm\infty}\mathcal{V}=
\left(-2ik^{2}+2i\beta^{2}q_{0}^{4}\right)\sigma_{3}+\left(2k+2\beta q_{0}^{2}\right)Q_{\pm},\label{LaxV2}
\end{align}\label{lax2}
\end{subequations}
where
\begin{eqnarray}
Q_{\pm}=\lim_{x\rightarrow\pm\infty}Q=\left(
\begin{array}{cc}
   0& q_{\pm} \\
 -\overline{q}_{\pm} & 0
\end{array}\right).
\nonumber\end{eqnarray}
it is obvious that  $\mathcal{V}_{\pm}$ and $\mathcal{U}_{\pm}$  possess  the  linear relationship
\begin{equation}
\mathcal{V}_{\pm}=\left(2k+2\beta q_{0}^{2}\right)\mathcal{U}_{\pm}.\nonumber
\end{equation}

The fundamental matrix solution of asymptotic spectral problem (\ref{lax2}) can be obtained as
\begin{equation}
\phi_{\pm}(x,t,k)=\Xi_{\pm}(k)e^{-i\theta(x,t,k)\sigma_{3}}, \quad\quad k\neq\beta q_{0}^{2}\pm iq_{0},\label{phipm}
\end{equation}
where
\begin{equation}
\theta(x,t,k)=\lambda\left[x+(2k+2\beta q_{0}^{2})t\right], \quad \lambda=\sqrt{(k-\beta q_{0}^{2})^{2}+q_{0}^{2}},\nonumber
\end{equation}
\begin{eqnarray}
\Xi_{\pm}=\left(
\begin{array}{cc}
   1& \frac{-iq_{\pm}}{ \lambda+(k-\beta q_{0}^{2})} \\
 \frac{-i\bar{q}_{\pm}}{\lambda+(k-\beta q_{0}^{2})} & 1
\end{array}\right).
\nonumber\end{eqnarray}

Since $\lambda$ is doubly branched with  branch points are $k=\beta q_{0}^{2}\pm iq_{0}$,  it is necessary to introduce a two-sheeted Riemann surface
to such  that $\lambda$  is a single-valued function on each sheet.

Denote $h=k-\beta q_{0}^{2}$ and let
\begin{equation}
h+iq_{0}=\rho_{1}e^{i\theta_{1}}, \quad h-iq_{0}=\rho_{2}e^{i\theta_{2}}, \ \ -\frac{\pi}{2}\leq\theta_{1},\theta_{2}\leq\frac{3\pi}{2},  \nonumber
\end{equation}
 one can rewrite  $\lambda$  on each sheet as
\begin{equation}
\lambda_{\mathrm{I}}=\sqrt{\rho_{1}\rho_{2}}e^{i\frac{\theta_{1}+\theta_{2}}{2}},
\quad \lambda_{\mathrm{II}}=-\lambda_{\mathrm{I}}=\sqrt{\rho_{1}\rho_{2}}e^{i(\frac{\theta_{1}+\theta_{2}}{2})+i\pi}.\nonumber
\end{equation}
And the branch cut of the Riemann surface is the segment $[-iq_{0},iq_{0}]$ in the complex $h$-plane.

Now we introduce a uniformization variable
\begin{equation}
z=\lambda+h,
\end{equation}
 then  its   inverse transformation  gives
\begin{equation}
\lambda=\frac{1}{2}\big(z+ q_{0}^2/z\big), \quad
h=\frac{1}{2}\big(z-q_{0}^2/z\big), \quad
k=\frac{1}{2}\big(z-q_{0}^2/z\big)+\beta q_{0}^{2}.
\end{equation}
Further  we can  show   the  following relations between the Riemann surface, the $h$-plane  and the $z$-plane.
\begin{itemize}
\item[-] {The region where $\mathrm{Im}\lambda>0$  come from  the upper-half plane of the sheet-$\mathrm{I}$ and
the lower-half plane of the sheet-$\mathrm{II}$.  The region where $\mathrm{Im}\lambda<0$  come from  the upper-half plane of the sheet-$\mathrm{II}$ and the lower-half plane of the sheet-$\mathrm{I}$.}
\item[-]{On the sheet-$\mathrm{I}$, $z\rightarrow\infty$ as $h\rightarrow\infty$, and on the sheet-$\mathrm{II}$, $z\rightarrow 0$ as $h\rightarrow\infty$.}
\item[-]{The real $\lambda$ (real $k$) axes is mapped into the real $z$ axes.}
\item[-]{The branch cut $[-iq_{0},iq_{0}]$ is mapped into the circle $C_{0}$ of the radius $q_{0}$ in $z$-plane.}
\item[-]{The sheet-$\mathrm{I}$ and sheet-$\mathrm{II}$, except for the branch cut, are mapped into the exterior and the interior of $C_{0}$, respectively.}
\end{itemize}
 The jump contour in the complex $z$-plane  is denoted by  $\Sigma=\mathbb{R}\cup C_{0}$.
The gray and white regions  in Fig.1  denote  $D^{+}$ and   $D^{-}$,  respectively
\begin{equation}
\begin{split}
&D^{+}=\{z\in \mathbb{C}|\mathrm{Im}\lambda=\left(|z|^{2}-q_{0}^{2}\right)\mathrm{Im}z>0\},\\
&D^{-}=\{z\in \mathbb{C}|\mathrm{Im}\lambda=\left(|z|^{2}-q_{0}^{2}\right)\mathrm{Im}z<0\}.
\end{split}\nonumber
\end{equation}
\begin{center}
{\includegraphics[scale=0.45]{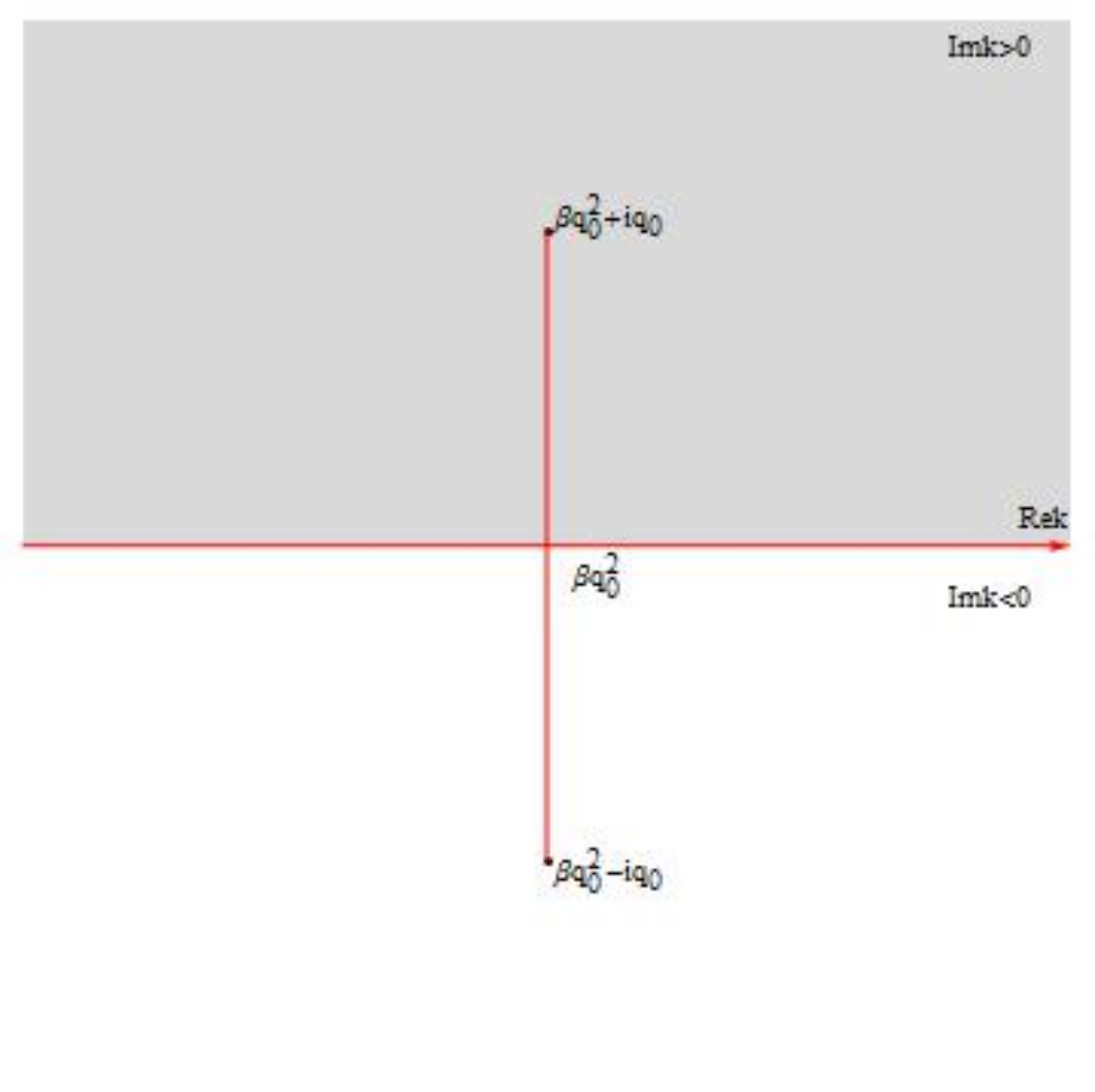}}\hspace{0.3cm}{\includegraphics[scale=0.40]{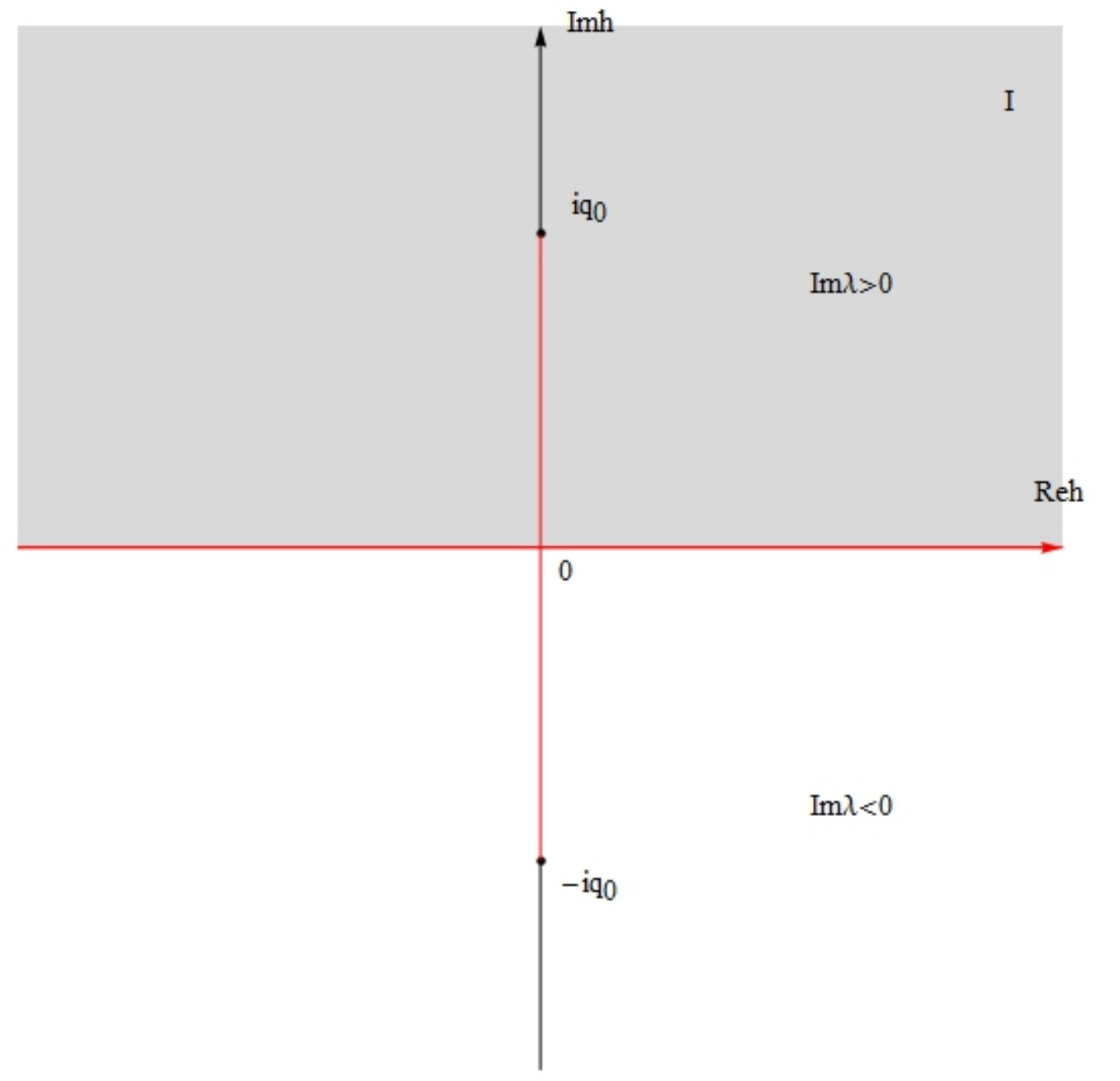}}\\
\vspace{-0.2cm}{\footnotesize\hspace{0.5cm}(a)\hspace{4.5cm}(b)}\\
{\includegraphics[scale=0.40]{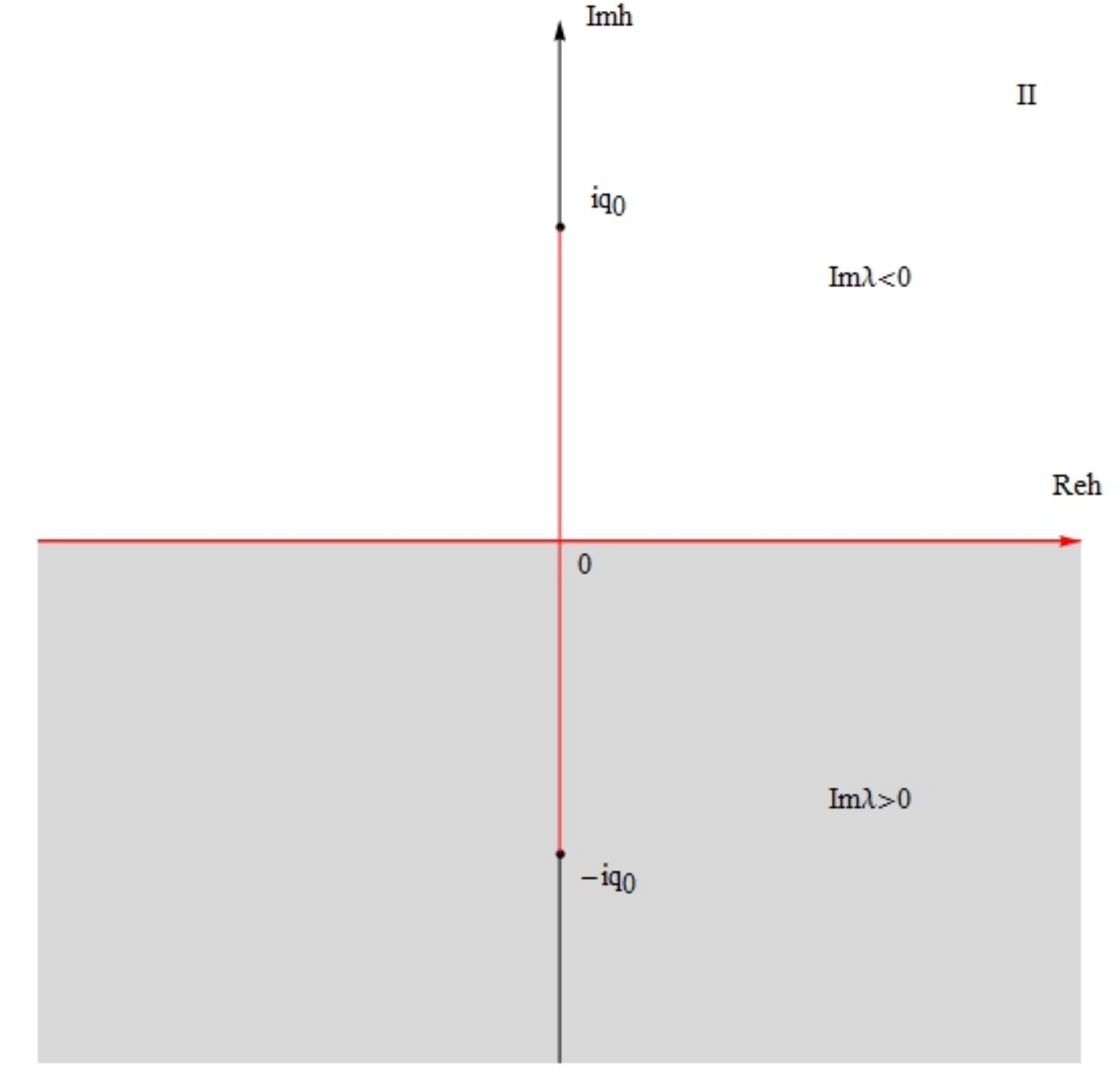}}\hspace{0.3cm}{\includegraphics[scale=0.40]{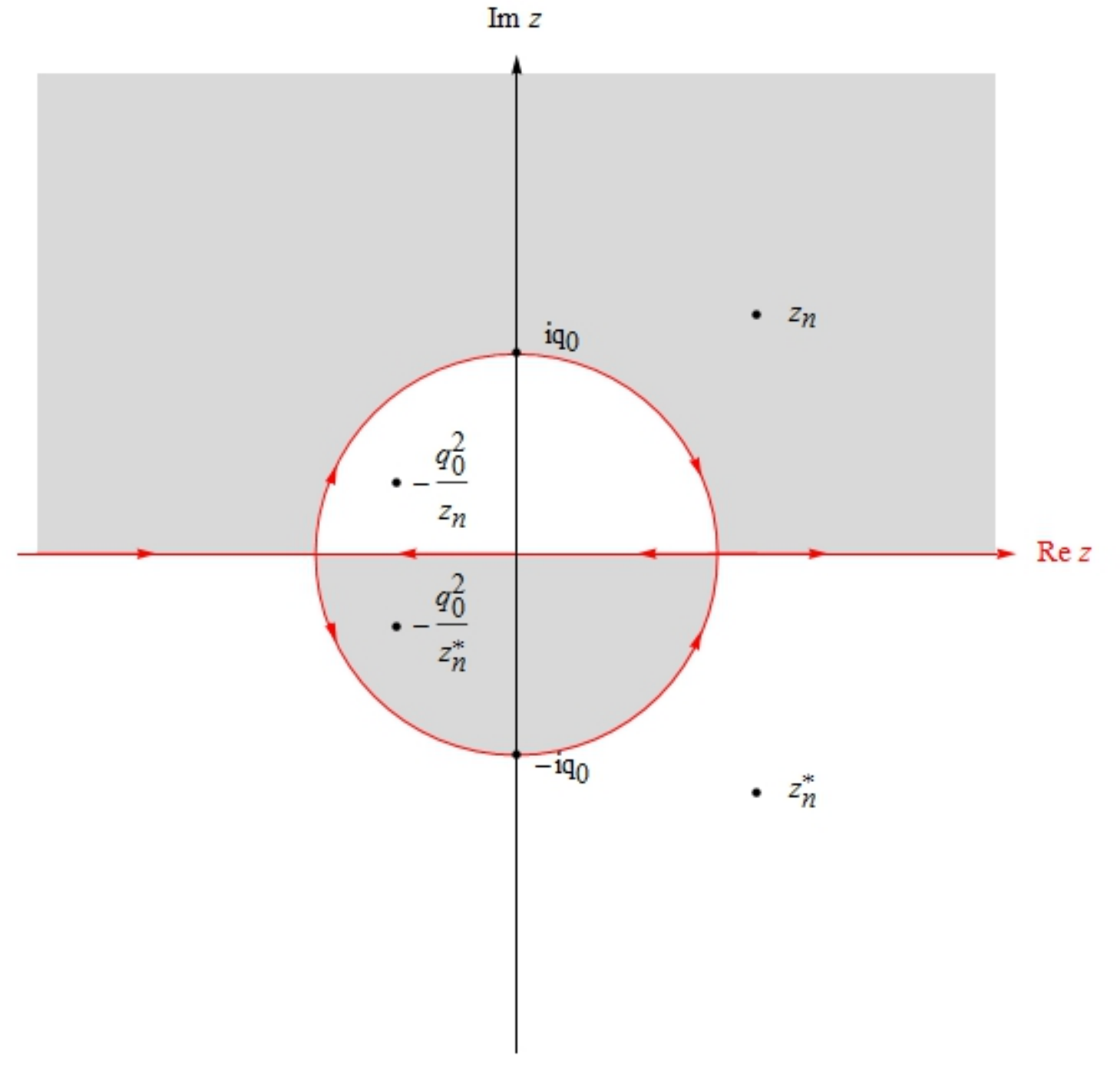}}\\
\vspace{-0.2cm}{\footnotesize\hspace{0.5cm}(c)\hspace{4.5cm}(d)}\\
 \flushleft{\footnotesize {\bf Fig.~1.} (a)The complex $k$-plane, the branch cut, $\mathrm{Im}k>0$(gray) and $\mathrm{Im}k<0$(white). (b)Sheet-$\mathrm{I}$ of Riemann surface, showing the branch cut (red) and the region where $\mathrm{Im}\lambda>0$(gray) and $\mathrm{Im}\lambda<0$(white). (c)Sheet-$\mathrm{II}$ of Riemann surface, showing the branch cut (red) and the region where $\mathrm{Im}\lambda>0$(gray) and $\mathrm{Im}\lambda<0$(white). (d)The complex $z$-plane, showing the region $D^{+}$ where $\mathrm{Im}\lambda>0$ (gray), the region $D^{-}$ where $\mathrm{Im}\lambda<0$(white), the orientation of the contours for the Riemann-Hilbert problem and the discrete spectrums(zero points of $s_{11}(z)$ and $s_{22}(z)$).}
\end{center}

Based on  the above results,  we can rewrite the fundamental matrix solution (\ref{phipm}) as
\begin{equation}
\phi_{\pm}(x,t,z)=\Xi_{\pm}(z)e^{-i\theta(x,t,z)\sigma_{3}}, \quad z\neq iq_{0}\nonumber
\end{equation}
 where
\begin{eqnarray}
\begin{split}
&\Xi_{\pm}(z)=\left(
\begin{array}{cc}
   1& \frac{-iq_{\pm}}{z} \\
 \frac{-i\bar{q}_{\pm}}{z} & 1
\end{array}\right)=\mathrm{I}-\frac{i}{z}\sigma_{3}Q_{\pm},
\nonumber\\
&\theta(x,t,z)=\frac{1}{2}\big(z+\frac{q_{0}^{2}}{z}\big)\left\{x+\Big[\big(z-\frac{q_{0}^{2}}{z}\big)+4\beta q_{0}^{2}\Big]t\right\}.\nonumber\\
\end{split}
\end{eqnarray}
Direct calculation shows  that
\begin{equation}
\begin{split}
&\det\Xi_{\pm}=1+\frac{q_{0}^{2}}{z^{2}}\triangleq\gamma\neq0, \  \  \
 \Xi_{\pm}^{-1}=\frac{1}{\gamma}\big(\mathrm{I}+\frac{i}{z}\sigma_{3}Q_{\pm}\big).\nonumber\\
\end{split}
\end{equation}

\section{Jost Solutions}
The Lax pair (\ref{Lax1}) can be rewrite as
\begin{subequations}
\begin{align}
&(\partial_{x}-\mathcal{U}_{\pm}-\Delta\hat{Q}_{\pm})\phi=0,\label{LaxU2}\\
&(\partial_{t}-\mathcal{V}_{\pm}-\Delta\hat{R}_{\pm})\phi=0,\label{LaxV2}
\end{align}\label{Lax2}
\end{subequations}
where
\begin{eqnarray}
\begin{split}
&\Delta\hat{Q}_{\pm}=i\beta\left(|q|^{2}-q_{0}^{2}\right)\sigma_{3}+\Delta Q_{\pm}, \quad \Delta\hat{R}_{\pm}=\hat{R}-\hat{R}_{\pm},\nonumber\\
&\Delta Q_{\pm}=Q-Q_{\pm}, \quad \hat{R}_{\pm}=2kQ_{\pm}+2\beta q_{0}^{2}Q_{\pm}+2i\beta^{2}q_{0}^{4}\sigma_{3},\nonumber\\
&\begin{split}\hat{R}=&2kQ+2\beta|q|^{2}Q+i\left(|q|^{2}+4\beta^{2}|q|^{4}-q_{0}^{2}-2\beta^{2}q_{0}^{4}\right)\sigma_{3}
        -iQ_{x}\sigma_{3}\\
        &-\beta\left(QQ_{x}-Q_{x}Q\right).\end{split}\nonumber
\end{split}
\end{eqnarray}
Now, one can define the Jost solutions as the simultaneous solutions of Lax pair (\ref{Lax1}) such that
\begin{equation}
\Psi_{\pm}(x,t,z)\sim\Xi_{\pm}(z)e^{-i\theta(x,t,z)\sigma_{3}}, \quad z\in\Sigma, \quad x\rightarrow\pm\infty,
\end{equation}
and the modified Jost solutions
\begin{eqnarray}
\begin{split}
e^{\int_{-\infty}^{x}i\beta\left(|q|^{2}-q_{0}^{2}\right) dy\hat{\sigma}_{3}}\mu_{\pm}(x,t,z)e^{-\int_{x}^{+\infty}i\beta\left(|q|^{2}-q_{0}^{2}\right) dy\sigma_{3}}=&\Psi_{\pm}(x,t,z)e^{i\theta(x,t,z)\sigma_{3}}\\
=&\eta_{\pm}(x,t,z),\label{mupsi}
\end{split}
\end{eqnarray}
Then the Lax pair (\ref{Lax2}) can be rewritten as
\begin{eqnarray}
\begin{split}
&\left(\Xi_{\pm}^{-1}\eta_{\pm}(x,t,z)\right)_{x}=-i\lambda\left[\sigma_{3},\Xi_{\pm}^{-1}\eta_{\pm}(x,t,z)\right]
+\Xi_{\pm}^{-1}\Delta\hat{Q}_{\pm}\eta_{\pm}(x,t,z),\\
&\begin{split}\left(\Xi_{\pm}^{-1}\eta_{\pm}(x,t,z)\right)_{t}
=&-i\lambda\left(2k+2\beta q_{0}^{2}\right)\left[\sigma_{3},\Xi_{\pm}^{-1}\eta_{\pm}(x,t,z)\right]\\
&+\Xi_{\pm}^{-1}\Delta\hat{R}_{\pm}\eta_{\pm}(x,t,z).\end{split}
\end{split}\label{Lax3}
\end{eqnarray}
It is easily known that Lax pair (\ref{Lax3}) can be written in full derivative form
\begin{equation}
d\left(e^{i\theta(x,t,z)\hat{\sigma}_{3}}\Xi_{\pm}^{-1}\eta_{\pm}(x,t,z)\right)
=e^{i\theta(x,t,z)\hat{\sigma}_{3}}\left(V_{1}dx+V_{2}dt\right),
\end{equation}
where
\begin{eqnarray}
\begin{split}
&V_{1}=\Xi_{\pm}^{-1}\Delta\hat{Q}_{\pm}\eta_{\pm}(x,t,z),\\
&V_{2}=\Xi_{\pm}^{-1}\Delta\hat{R}_{\pm}\eta_{\pm}(x,t,z).
\end{split}\nonumber
\end{eqnarray}
We can obtain the Volterra integral equations
\begin{eqnarray}
\begin{split}
&\begin{split}\mu_{-}(x,t,z)=\Xi_{-}+\int_{-\infty}^{x}&e^{-\int_{-\infty}^{x}i\beta\left(|q|^{2}-q_{0}^{2}\right) dy\sigma_{3}}\Xi_{-}e^{i\lambda(x'-x)\hat{\sigma}_{3}}\\
&\Xi_{-}^{-1}\Delta\hat{Q}_{-}e^{\int_{-\infty}^{x'}i\beta\left(|q|^{2}-q_{0}^{2}\right) dy\sigma_{3}}\mu_{-}dx',
\end{split}\\
&\begin{split}\mu_{+}(x,t,z)=\Xi_{+}-\int_{x}^{+\infty}&e^{-\int_{-\infty}^{x}i\beta\left(|q|^{2}-q_{0}^{2}\right) dy\sigma_{3}}\Xi_{+}e^{i\lambda(x'-x)\hat{\sigma}_{3}}\\
&\Xi_{+}^{-1}\Delta\hat{Q}_{+}e^{\int_{-\infty}^{x'}i\beta\left(|q|^{2}-q_{0}^{2}\right) dy\sigma_{3}}\mu_{+}dx'.\end{split}
\end{split}
\end{eqnarray}

It  can be shown that  the first column of $\mu_{-}$ is  analytically extended to $D^{+}$ and  the second column of $\mu_{-}$  is  analytically extended to $D^{-}$.
Similarly, the first column of $\mu_{+}$ can be analytically extended to $D^{-}$ and the second column of $\mu_{+}$ can be analytically extended to $D^{+}$. It can be summarized as follows
\begin{equation}
\begin{split}
D^{+}: \quad \mu_{-,1}, \quad \mu_{+,2};\\
D^{-}: \quad \mu_{-,2}, \quad \mu_{+,1},
\end{split}\nonumber
\end{equation}
where the subscripts `1' and `2' identify the columns of matrix.

Consider the Laurent expansion of $\Xi_{\pm}^{-1}\eta_{\pm}(x,t,z)$  in  system (\ref{Lax3})
\begin{subequations}
\begin{align}
&\Xi_{\pm}^{-1}\eta_{\pm}(x,t,z)=\alpha^{(0)}+\frac{\alpha^{(1)}}{z}+\frac{\alpha^{(2)}}{z^{2}}+O\big(\frac{1}{z^{3}}\big),\quad\quad z\rightarrow\infty,\\
&\Xi_{\pm}^{-1}\eta_{\pm}(x,t,z)=\eta^{(0)}+\eta^{(1)}z+\eta^{(2)}z^{2}+O\left(z^{3}\right), \quad z\rightarrow 0.
\end{align}
\end{subequations}
where $\alpha^{(0)},\alpha^{(1)},\alpha^{(2)},\cdots$ and $\eta^{(0)},\eta^{(1)},\eta^{(2)},\cdots$are independent of $z$.
Substituting the above expansion into (\ref{Lax3}) and comparing the coefficients of $z^{n} (n=0,\pm1,\pm2,\cdots)$, we obtain the following equations
\begin{eqnarray}
\begin{split}
&\left[\sigma_{3},\alpha^{(0)}\right]=0,\nonumber\\
&\alpha^{(0)}_{x}=-\frac{i}{2} \left[\sigma_{3},\alpha^{(1)}\right]+i\beta\left(|q|^{2}-q_{0}^{2}\right)\sigma_{3}\alpha^{(0)}+\Delta Q_{\pm}\alpha^{(0)},\nonumber\\
&\begin{split}\alpha^{(1)}_{x}=&-\frac{i}{2}q_{0}^{2} \left[\sigma_{3},\alpha^{(0)}\right]-\frac{i}{2}\left[\sigma_{3},\alpha^{(2)}\right]
+\beta\left(|q|^{2}-q_{0}^{2}\right)\left[\sigma_{3},\sigma_{3}Q_{\pm}\right]\alpha^{(0)}\\
&+i\left[\sigma_{3}Q_{\pm},\Delta Q_{\pm}\right]\alpha^{(0)}+i\beta\left(|q|^{2}-q_{0}^{2}\right)\sigma_{3}\alpha^{(1)}+\Delta Q_{\pm}\alpha^{(1)},
\end{split}\nonumber
\end{split}
\end{eqnarray}
\begin{eqnarray}
\begin{split}
&\left[\sigma_{3},\eta^{(0)}\right]=0,\nonumber\\
&\eta^{(0)}_{x}=-\frac{i}{2}q_{0}^{2} \left[\sigma_{3},\eta^{(1)}\right]+\frac{1}{q_{0}^{2}}i\beta\left(|q|^{2}-q_{0}^{2}\right)\sigma_{3}\eta^{(0)}+\frac{1}{q_{0}^{2}}\sigma_{3}Q_{\pm}\Delta Q_{\pm}\sigma_{3}Q_{\pm}\eta^{(0)},\nonumber\\
&\begin{split}\eta^{(1)}_{x}=&-\frac{i}{2} \left[\sigma_{3},\eta^{(0)}\right]-\frac{i}{2}q_{0}^{2}\left[\sigma_{3},\eta^{(2)}\right]
+\frac{1}{q_{0}^{2}}\beta\left(|q|^{2}-q_{0}^{2}\right)\left[\sigma_{3},\sigma_{3}Q_{\pm}\right]\eta^{(0)}\\
&+\frac{i}{q_{0}^{2}}\left[\sigma_{3}Q_{\pm},\Delta Q_{\pm}\right]\eta^{(0)}+\frac{1}{q_{0}^{2}}i\beta\left(|q|^{2}-q_{0}^{2}\right)\sigma_{3}Q_{\pm}^{2}\eta^{(1)}\\
&+\frac{1}{q_{0}^{2}}\sigma_{3}Q_{\pm}\Delta Q_{\pm}\sigma_{3}Q_{\pm}\eta^{(1)},
\end{split}\nonumber
\end{split}
\end{eqnarray}
from which,  we can derive the following results
\begin{eqnarray}
\begin{split}
&\alpha^{(0)}=e^{-\int_{x}^{+\infty}i\beta\left(|q|^{2}-q_{0}^{2}\right)dy\sigma_{3}},\\
&\alpha^{(1)}_{o}=-i\sigma_{3}\Delta Q_{\pm}\alpha^{(0)}_{d},\\
&\alpha^{(1)}_{d,x}=i\left[\sigma_{3}Q_{\pm},\Delta Q_{\pm}\right]\alpha^{(0)}_{d}+i\beta\left(|q|^{2}-q_{0}^{2}\right)\sigma_{3}\alpha^{(1)}_{d}-i\Delta Q_{\pm}\sigma_{3}\Delta Q_{\pm}\alpha^{(0)}_{d}.
\end{split}\nonumber
\end{eqnarray}
and
\begin{eqnarray}
\begin{split}
&\eta^{(0)}=e^{\int_{x}^{+\infty}i\beta\left(|q|^{2}-q_{0}^{2}\right)dy\sigma_{3}},\\
&\begin{split}\eta^{(1)}_{d,x}=&\frac{i}{q_{0}^{2}}\left[\sigma_{3}Q_{\pm},\Delta Q_{\pm}\right]\eta^{(0)}_{d}-i\beta\left(|q|^{2}-q_{0}^{2}\right)\sigma_{3}\eta^{(1)}_{d}\\
&+\frac{1}{q_{0}^{4}}\sigma_{3}Q_{\pm}\Delta Q_{\pm}\sigma_{3}\Delta Q_{\pm}\sigma_{3}Q_{\pm}\eta^{(0)}_{d}.
\end{split}
\end{split}\nonumber
\end{eqnarray}
And the asymptotic of the modified Jost solutions for $z\rightarrow\infty$ and $z\rightarrow 0$ are respectively derived as
\begin{eqnarray}
\begin{split}
&\begin{split}\mu_{\pm}=&I-\frac{i}{z}e^{-\int_{-\infty}^{x}i\beta\left(|q|^{2}-q_{0}^{2}\right)dy\hat{\sigma}_{3}}\sigma_{3}Q_{\pm}\\
&+\frac{1}{z}e^{-\int_{-\infty}^{x}i\beta\left(|q|^{2}-q_{0}^{2}\right)dy\hat{\sigma}_{3}}\alpha^{(1)}e^{\int_{x}^{+\infty}i\beta\left(|q|^{2}-q_{0}^{2}\right)dy\sigma_{3}}+O\big(\frac{1}{z^{2}}\big), \quad z\rightarrow\infty,\end{split}\\
&\mu_{\pm}=-\frac{i}{z}\sigma_{3}Q_{\pm}+O\left(1\right), \quad z\rightarrow 0.
\end{split}\label{muexp}
\end{eqnarray}

\section{Scattering Matrix and Asymptotic}

It is easy to check that  $\mathrm{tr}\mathcal{U}=\mathrm{tr}\mathcal{V}=0$,  then by using Abel formula,
we have
$$(\det\Psi_{\pm})_{x}=(\det\Psi_{\pm})_{t}=0,$$
which implies that
\begin{align}
&\det\Psi_{\pm}=\det Y_{\pm}=\gamma\label{ppo}
\end{align}
 by using asymptotic at $x\rightarrow \pm \infty$.
 Since  $\Psi_{+}$ and $\Psi_{-}$ are the fundamental solutions of the spectral problem (\ref{Lax1}),  they satisfy the following linear relationship
\begin{equation}
\Psi_{+}(z)=\Psi_{-}(z)S(z), \quad z\in\Sigma\setminus\{\pm iq_{0}\},\label{S}
\end{equation}
where $S(z)=(s_{ij}(z))_{2\times 2}$ is called spectral matrix.   From (\ref{ppo}), we know that    $\det S=1$.
The relation formula  (\ref{S}) can be expanded as
\begin{eqnarray}
\Psi_{+,1}=s_{11}\Psi_{-,1}+s_{21}\Psi_{-,2}, \quad
\Psi_{+,2}=s_{12}\Psi_{-,1}+s_{22}\Psi_{-,2}.\label{psis}
\end{eqnarray}
The reflection coefficients are defined as
\begin{equation}
\rho(z)=\frac{s_{21}}{s_{11}}, \quad \tilde{\rho}(z)=\frac{s_{12}}{s_{22}}.\label{rho}
\end{equation}
According to (\ref{psis}) the scattering coefficients have the following Wronskian representations
\begin{subequations}
\begin{align}
&s_{11}=\frac{\mathrm{Wr}\left(\Psi_{+,1},\Psi_{-,2}\right)}{\gamma}, \quad s_{12}=\frac{\mathrm{Wr}\left(\Psi_{+,2},\Psi_{-,2}\right)}{\gamma},\\
&s_{21}=\frac{\mathrm{Wr}\left(\Psi_{-,1},\Psi_{+,1}\right)}{\gamma}, \quad s_{22}=\frac{\mathrm{Wr}\left(\Psi_{-,1},\Psi_{+,2}\right)}{\gamma}.
\end{align}\label{sij}
\end{subequations}
 The equation (\ref{sij}) implies that $s_{11}$ is analytic in $D^{-}$ and $s_{22}$ is analytic in $D^{+}$.
  However, $s_{12}$ and $s_{22}$  are just continuous   on $\Sigma$.

\noindent\textbf{Proposition 1.} The asymptotic behaviors of the scattering matrix $S(z)$ are given as follows
\begin{subequations}
\begin{align}
&S(z)=I+O\big(\frac{1}{z}\big),\quad z\rightarrow\infty,\label{s1}\\
&S(z)=\mathrm{diag}\left( q_{-}/q_{+}, q_{+}/q_{-}\right)+O\left(z\right),\quad z\rightarrow 0.\label{s2}
\end{align}\label{s}
\end{subequations}
\begin{proof}
By using  (\ref{mupsi}), (\ref{sij}) and the asymptotic behaviors of $\mu_{\pm}$,  we can prove that

As $z\rightarrow\infty$,
\begin{eqnarray}
\begin{split}
s_{11}=&\frac{Wr\left(\Psi_{+,1},\Psi_{-,2}\right)}{\gamma} = \frac{1 }{1+\frac{q_{0}^{2}}{z^{2}}}  \det
\left(
\begin{array}{cc}
   \mu_{+,11}& \mu_{-,12}\\
\mu_{+,21} & \mu_{-,22}
\end{array}\right)\nonumber\\
=&\det
\left(
\begin{array}{cc}
   1+O(\frac{1}{z})& O(\frac{1}{z})\\
O(\frac{1}{z}) & 1+O(\frac{1}{z})
\end{array}\right)\big(1-\frac{q_{0}^{2}}{z^{2}}+\frac{q_{0}^{4}}{z^{4}}-\cdots\big)\\
=&\big(1+O\big(\frac{1}{z}\big)\big)\big(1-\frac{q_{0}^{2}}{z^{2}}+\cdots\big)\\
=&1+O\big(\frac{1}{z}\big).
\end{split}
\end{eqnarray}

As $z\rightarrow 0$,
\begin{eqnarray}
\begin{split}
s_{11}=&\frac{Wr\left(\Psi_{+,1},\Psi_{-,2}\right)}{\gamma}=\frac{ 1}{1+\frac{q_{0}^{2}}{z^{2}}}  \det
\left(
\begin{array}{cc}
   \mu_{+,11}& \mu_{-,12} \\
\mu_{+,21} & \mu_{-,22}
\end{array}\right)\nonumber\\
=&\det
\left(
\begin{array}{cc}
   O(1)& -\frac{i}{z}q_{-}\\
\frac{i}{z}\overline{q}_{+} & O(1)
\end{array}\right)\frac{z^{2}}{q_{0}^{2}}\big(1-\frac{z^{2}}{q_{0}^{2}}+\cdots\big)\\
=&\big(O(1)+\frac{1}{z^{2}}q_{-}\overline{q}_{+}\big)\big(\frac{z^{2}}{q_{0}^{2}}-\frac{z^{4}}{q_{0}^{4}}+\cdots\big)\\
=&\frac{q_{-}\overline{q}_{+}}{q_{0}^{2}}+O\left(z\right)=\frac{q_{-}}{q_{+}}+O(z).
\end{split}
\end{eqnarray}
The asymptotic behaviors of $s_{22}$, $s_{12}$, and $s_{21}$ can also be derived by the similar way.
\end{proof}

\section{Symmetries}

For  the focusing KE equation with nonzero boundary conditions,   the  Jost functions  $\Psi_{\pm}(z)$  and spectral matrix  $S(z)$
  possess two kinds of symmetries.

\subsection{First Symmetry}

Here we consider the   symmetries between two points $z\mapsto \overline{z}$ (upper/lower half plane).\\
\noindent\textbf{Proposition 2.} For $z\in \Sigma$,\\
(1) The Jost solutions satisfy the symmetries
\begin{subequations}
\begin{align}
&\Psi_{\pm}(z)=\sigma_{2}\overline{\Psi_{\pm}(\overline{z})}\sigma_{2},\\ &\Psi_{\pm,1}(z)=i\sigma_{2}\overline{\Psi_{\pm,2}(\overline{z})}, \\ &\Psi_{\pm,2}=-i\sigma_{2}\overline{\Psi_{\pm}(\overline{z})}.
\end{align}\label{psisym1}
\end{subequations}
(2) The scattering matrix satisfy the symmetries
\begin{subequations}
\begin{align}
&S(z)=\sigma_{2}\overline{S(\overline{z})}\sigma_{2},\\
&s_{11}(z)=\overline{s_{22}(\overline{z})}, \\
&s_{12}(z)=-\overline{s_{21}(\overline{z})}.\label{ssym1}
\end{align}
\end{subequations}
(3) The reflection coefficient satisfy the symmetries
\begin{align}
\rho(z)=-\overline{\tilde{\rho}(\overline{z})}.\label{rhosym1}
\end{align}

\begin{proof}
(1) The $\mathcal{U}$ and $\mathcal{V}$ in the Lax pair (\ref{Lax1}) with the following symmetries on $z$-plane
\begin{equation}
\overline{\mathcal{U}(\overline{z})}=\sigma_{2}\mathcal{U}(z)\sigma_{2}, \quad \overline{\mathcal{V}(\overline{z})}=\sigma_{2}\mathcal{V}(z)\sigma_{2},\nonumber
\end{equation}
by which, we can show that
\begin{eqnarray}
\begin{split}
&\big(\sigma_{2}\overline{\Psi_{\pm}(\overline{z})}\sigma_{2}\big)_{x}=\mathcal{U}(z)\big(\sigma_{2}\overline{\Psi_{\pm}(\overline{z})}\sigma_{2}\big),\\
&\big(\sigma_{2}\overline{\Psi_{\pm}(\overline{z})}\sigma_{2}\big)_{t}=\mathcal{V}(z)\big(\sigma_{2}\overline{\Psi_{\pm}(\overline{z})}\sigma_{2}\big),
\end{split}\nonumber
\end{eqnarray}
Further, by suing asymptotic
\begin{equation}
\Psi_{\pm}(z), \ \sigma_{2}\overline{\Psi_{\pm}(\overline{z})}\sigma_{2}\sim\Xi_{\pm}(z)e^{-i\lambda\left[x+(2k+2\beta q_{0}^{2})t\right]\sigma_{3}}, \ \ x\rightarrow \infty,\nonumber
\end{equation}
 we  obtain the symmetry  (\ref{psisym1}).\\
(2) On the basis of (\ref{S}),
\begin{eqnarray}
\begin{split}
S(z)=&\Psi_{-}^{-1}(z)\Psi_{+}(z)\\
    =&\sigma_{2}\overline{\Psi_{-}(\overline{z})}^{-1}\sigma_{2}\sigma_{2}\overline{\Psi_{+}(\overline{z})}\sigma_{2}\\
    =&\sigma_{2}\overline{\Psi_{-}(\overline{z})}^{-1}\overline{\Psi_{+}(\overline{z})}\sigma_{2}\\
    =&\sigma_{2}\overline{S(\overline{z})}\sigma_{2}.
\end{split}\nonumber
\end{eqnarray}
Then (\ref{ssym1}) can be obtained.\\
(3) \begin{equation}
\rho(z)=\frac{s_{21}(z)}{s_{11}(z)}=-\frac{\overline{s_{12}(\overline{z})}}{\overline{s_{22}(\overline{z})}}=-\overline{\tilde{\rho}(\overline{z})}.\nonumber
\end{equation}
\end{proof}

\subsection{Second Symmetry}

Here  we consider the   symmetries between two points  $z\mapsto -\frac{q_{0}^{2}}{z}$ (outside/inside of the circle $C_{0})$.\\
\noindent\textbf{Proposition 3.} For $z\in \Sigma$,\\
(1) The Jost solutions satisfy the following symmetries
\begin{subequations}
\begin{align}
&\Psi_{\pm}(z)=-\frac{i}{z}\Psi_{\pm}\big(-\frac{q_{0}^{2}}{z}\big)\sigma_{3}Q_{\pm},\\ &\Psi_{\pm,1}(z)=-\frac{i}{z}\overline{q}_{\pm}\Psi_{\pm,2}\big(-\frac{q_{0}^{2}}{z}\big), \\ &\Psi_{\pm,2}(z)=-\frac{i}{z}q_{\pm}\Psi_{\pm,1}\big(-\frac{q_{0}^{2}}{z}\big),
\end{align}\label{psisym2}
\end{subequations}
(2) The scattering matrix satisfies the symmetries
\begin{subequations}
\begin{align}
&S(z)=\left(\sigma_{3}Q_{-}\right)^{-1}S\big(-\frac{q_{0}^{2}}{z}\big)\sigma_{3}Q_{+}, \\
&s_{11}\big(-\frac{q_{0}^{2}}{z}\big)=\frac{q_{-}}{q_{+}}s_{22}(z), \\
&s_{12}\big(-\frac{q_{0}^{2}}{z}\big)=\frac{q_{+}}{\overline{q}_{-}}s_{21}(z).
\end{align} \label{ssym1}
\end{subequations}
(3) The reflection coefficient satisfies the symmetries
\begin{equation}
\rho(-\frac{q_{0}^{2}}{z})=-\frac{\overline{q}_{-}}{q_{-}}\overline{\rho(\overline{z})}.\label{rhosym1}
\end{equation}
\begin{proof}
The process of proof  is similar with the Proposition 2.
\end{proof}

\section{Riemann-Hilbert Problem}

Based on the analytical and asymmetry properties of  eigenfunctions  $\mu_\pm$ and $S(z)$,  we  derive Riemann-Hilbert problem associated with
KE equation with nonzero boundary conditions. \\

\noindent\textbf{Proposition 4.}   Define   sectionally meromorphic matrix
\begin{eqnarray}
M(x,t,z)=\begin{cases}
\begin{split}
&M^{+}(x,t,z)=\left(
\begin{array}{cc}
  \mu_{-,1}& \frac{\mu_{+,2}}{s_{22}}
\end{array}\right), \quad z\in D^{+},\\
&M^{-}(x,t,z)=\left(
\begin{array}{cc}
  \frac{\mu_{+,1}}{s_{11}}& \mu_{-,2}
\end{array}\right), \quad z\in D^{-},
\end{split}
\end{cases}\label{M}
\end{eqnarray}
then we have the following  Riemann-Hilbert problem
\begin{itemize}
\item[-] {Analyticity: $M(x,t,z)$ is  meromorphic  in $D^{+}\cup D^{-}$ and has simple poles.}
\item[-] {Jump condition:
\begin{equation}
M^{-}(x,t,z)=M^{+}(x,t,z)\left(I-G(x,t,z)\right), \quad z\in\Sigma,\nonumber
\end{equation}
where
\begin{eqnarray}
G(x,t,z)=e^{-i\theta\sigma_{3}}\left(
\begin{array}{cc}
   \rho\tilde{\rho}& \tilde{\rho} \\
-\rho & 0
\end{array}\right)e^{i\theta\sigma_{3}}.\nonumber
\end{eqnarray}}
\item[-] {Asymptotic behavior:
\begin{eqnarray}
\begin{split}
&M^{\pm}(x,t,z)=I+O\big(\frac{1}{z}\big),\quad z\rightarrow\infty\\
&M^{\pm}(x,t,z)=-\frac{i}{z}\sigma_{3}Q_{-}+O(1).\quad z\rightarrow 0
\end{split}\nonumber
\end{eqnarray}}
\end{itemize}
\begin{proof}
The analyticity can be derived from (\ref{psis}) and the analyticity of $\mu_{\pm}$.

From (\ref{psis}), we can known that
\begin{eqnarray}
\begin{split}
&\frac{\Psi_{+,1}}{s_{11}}=\left(1-\rho\tilde{\rho}\right)\Psi_{-,1}+\rho\frac{\Psi_{+,2}}{s_{22}},\\
&\Psi_{-,2}=-\tilde{\rho}\Psi_{-,1}+\frac{\Psi_{+,2}}{s_{22}},
\end{split}\nonumber
\end{eqnarray}
which lead to
\begin{eqnarray}
\left(
\begin{array}{cc}
   \frac{\Psi_{+,1}}{s_{11}}& \Psi_{-,2}
\end{array}\right)=\left(
\begin{array}{cc}
   \Psi_{-,1}&\frac{\Psi_{+,2}}{s_{22}}
\end{array}\right)\left(
\begin{array}{cc}
   1-\rho\tilde{\rho}&-\tilde{\rho}\\
   \rho&1
\end{array}\right).\nonumber
\end{eqnarray}
Then the jump condition can be derived as
\begin{eqnarray}
\begin{split}M^{-}(x,t,z)=&\left(
\begin{array}{cc}
   \frac{\mu_{+,1}}{s_{11}}& \mu_{-,2}
\end{array}\right)\\
=&\left(
\begin{array}{cc}
   \mu_{-,1}&\frac{\mu_{+,2}}{s_{22}}
\end{array}\right)e^{-i\theta\sigma_{3}}\left(
\begin{array}{cc}
   1-\rho\tilde{\rho}&-\tilde{\rho}\\
   \rho&1
\end{array}\right)e^{i\theta\sigma_{3}}\\
=&M^{+}(x,t,z)\left(I-G(x,t,z)\right).
\end{split}\nonumber
\end{eqnarray}
Now we proof the asymptotic behavior. From (\ref{M}), $M^{+}(x,t,z)$ with the following form
\begin{eqnarray}
\begin{split}
M^{+}(x,t,z)=&\left(
\begin{array}{cc}
   \mu_{-,1}&\mu_{+,2}
\end{array}\right)\left(
\begin{array}{cc}
   1&0\\
   0&\frac{1}{s_{22}}
\end{array}\right).
\end{split}\nonumber
\end{eqnarray}
As $z\rightarrow\infty$,
\begin{eqnarray}
\begin{split}
M^{+}(x,t,z)=&\left(I+O\big(\frac{1}{z}\big)\right)\left(
\begin{array}{cc}
   1&0\\
   0&\frac{1}{1+O\left(\frac{1}{z}\right)}
\end{array}\right)\\
=&I+O\big(\frac{1}{z}\big).
\end{split}\nonumber
\end{eqnarray}
As $z\rightarrow 0$,
\begin{eqnarray}
\begin{split}
M^{+}(x,t,z)=&\left(
\begin{array}{cc}
   O(1)&-\frac{i}{z}q_{+}+O(1)\\
   -\frac{i}{z}\overline{q}_{-}+O(1)&O(1)
\end{array}\right)\left(
\begin{array}{cc}
   1&0\\
   0&\frac{1}{\frac{q_{+}}{q_{-}}+O(z)}
\end{array}\right)\\
=&-\frac{i}{z}\sigma_{3}Q_{-}+O(1).
\end{split}\nonumber
\end{eqnarray}
The asymptotic behavior of $M^{-}(x,t,z)$ can be derived in the similar way.
\end{proof}

\section{Discrete Spectrum and Residue Conditions}
If $s_{22}(z_{n})=0$, we can derive from the forth equation of (\ref{sij}) that the eigenfunctions $\Psi_{+,2}(x,t,z_{n})$ and $\Psi_{-,1}(x,t,z_{n})$ must be proportional
\begin{equation}
\Psi_{+,2}(x,t,z_{n})=b_{n}\Psi_{-,1}(x,t,z_{n}),\label{psib}
\end{equation}
where $b_{n}$ is independent of $x$, $t$ and $z$. Let $s_{22}$ has a finite number of simple zeros $z_{1}, z_{2},\cdots, z_{N}$ in $D^{+}\cap\{z\in\mathbb{C}|\mathrm{Im}z>0\}$. According to the symmetry properties of $S(z)$, we have that
\begin{equation}
s_{22}(z_{n})=0\Longleftrightarrow s_{11}(\overline{z}_{n})=0\Longleftrightarrow s_{11}(-\frac{q_{0}^{2}}{z_{n}})=0\Longleftrightarrow s_{22}(-\frac{q_{0}^{2}}{\overline{z}_{n}})=0.\nonumber
\end{equation}
So the discrete spectrum accumulate the set
\begin{equation}
\mathcal{Z}=\{z_{n},\overline{z}_{n},-\frac{q_{0}^{2}}{z_{n}},-\frac{q_{0}^{2}}{\overline{z}_{n}}\}_{n=1}^{N}.\nonumber
\end{equation}

Now we study the residue condition that is very important to  solve  the Riemann-Hilbert problem. The equation (\ref{psib}) can be rewritten as
\begin{equation}
\mu_{+,2}(z_{n})=b_{n}\mu_{-,1}(z_{n})e^{-2i\theta(z_{n})}.\nonumber
\end{equation}
Then
\begin{eqnarray}
\begin{split}
\mathop{\mathrm{Res}}\limits_{z=z_{n}}\left[\frac{\mu_{+,2}(z)}{s_{22}(z)}\right]=&\frac{b_{n}\mu_{-,1}(z_{n})e^{-2i\theta(z_{n})}}{s_{22}^{'}(z_{n})}=A_{n}\mu_{-,1}(z_{n})e^{-2i\theta(z_{n})},
\end{split}
\end{eqnarray}
where $A_{n}=\frac{b_{n}}{s_{22}^{'}(z_{n})}$.

If $s_{11}(\overline{z}_{n})=0$, similarly, the first equation of (\ref{sij}) implies that the eigenfunctions $\Psi_{+,1}(x,t,\overline{z}_{n})$ and $\Psi_{-,2}(x,t,\overline{z}_{n})$ must be proportional
\begin{equation}
\Psi_{+,1}(x,t,\overline{z}_{n})=\tilde{b}_{n}\Psi_{-,2}(x,t,\overline{z}_{n}),\label{psib1}
\end{equation}
(\ref{psib1}) also can be rewritten as
\begin{equation}
\mu_{+,1}(\overline{z}_{n})=\tilde{b}_{n}\mu_{-,2}(\overline{z}_{n})e^{2i\theta(\overline{z}_{n})}.\nonumber
\end{equation}
And derived that
\begin{eqnarray}
\begin{split}
\mathop{\mathrm{Res}}\limits_{z=\overline{z}_{n}}\left[\frac{\mu_{+,1}(z)}{s_{11}(z)}\right]=&\frac{\tilde{b}_{n}\mu_{-,2}
(\overline{z}_{n})e^{-2i\theta(\overline{z}_{n})}}{s_{11}^{'}(\overline{z}_{n})}=\tilde{A}_{n}\mu_{-,2}(\overline{z}_{n})e^{2i\theta(\overline{z}_{n})},
\end{split}
\end{eqnarray}
where $\tilde{A}_{n}=\frac{\tilde{b}_{n}}{s_{11}^{'}(\overline{z}_{n})}$.
And it is easy to show that
\begin{equation}
\tilde{A}_{n}=-\overline{A}_{n}.
\end{equation}
Combining (\ref{psisym2}), (\ref{psib}) and (\ref{psib1}) we have the following relations
\begin{subequations}
\begin{align}
&\Psi_{+,2}\big(-\frac{q_{0}^{2}}{\overline{z}_{n}}\big)=\frac{q_{-}}{\overline{q}_{+}}\tilde{b}_{n}\Psi_{-,1}\big(-\frac{q_{0}^{2}}{\overline{z}_{n}}\big),\\
&\Psi_{+,1}\big(-\frac{q_{0}^{2}}{z_{n}}\big)=\frac{\overline{q}_{-}}{q_{+}}b_{n}\Psi_{-,2}\big(-\frac{q_{0}^{2}}{z_{n}}\big).
\end{align}
\end{subequations}
Using the symmetries of $s_{ij}$, we have results
\begin{subequations}
\begin{align}
&\mathop{\mathrm{Res}}\limits_{z=-\frac{q_{0}^{2}}{\overline{z}_{n}}}\left[\frac{\mu_{+,2}(z)}{s_{22}(z)}\right]
=A_{N+n}\mu_{-,1}(-\frac{q_{0}^{2}}{\overline{z}_{n}})e^{-2i\theta(-\frac{q_{0}^{2}}{\overline{z}_{n}})},\\
&\mathop{\mathrm{Res}}\limits_{z=-\frac{q_{0}^{2}}{z_{n}}}\left[\frac{\mu_{+,1}(z)}{s_{11}(z)}\right]
=\tilde{A}_{N+n}\mu_{-,2}(-\frac{q_{0}^{2}}{z_{n}})e^{2i\theta(-\frac{q_{0}^{2}}{z_{n}})}.
\end{align}
\end{subequations}
where
\begin{equation}
A_{N+n}=\frac{q_{-}}{\overline{q}_{-}}\frac{q_{0}^{2}}{\overline{z}_{n}^{2}}\tilde{A}_{n}, \quad \tilde{A}_{N+n}=\frac{\overline{q}_{-}}{q_{-}}\frac{q_{0}^{2}}{z_{n}^{2}}A_{n},
\end{equation}
and $\tilde{A}_{N+n}=-\overline{A}_{N+n}$.

\section{N-soliton solutions  of the KE equation}

To obtain solution of KE equation with nonzero boundary conditions, we should
establish the connection between  the solution of KE equation and the Riemann-Hilbert problem.

\subsection{Reconstruction formula}

To solve the above Riemann-Hilbert problem, it is necessary to regularize it by subtracting out the asymptotic and the pole contributions.
For convenient, we define $\zeta_{n}=z_{n}$ and $\zeta_{N+n}=-\frac{q_{0}^{2}}{\overline{z}_{n}}$, $(n=1,\cdots,N)$ and
\begin{eqnarray}
\begin{split}
&M^{-}-I+\frac{i}{z}\sigma_{3}Q_{-}-\mathop{\mathrm{\sum}}\limits_{n=1}^{2N}\frac{\mathop{\mathrm{Res}}\limits_{\overline{\zeta}_{n}}M^{-}}{z-\overline{\zeta}_{n}}
-\mathop{\mathrm{\sum}}\limits_{n=1}^{2N}\frac{\mathop{\mathrm{Res}}\limits_{\zeta_{n}}M^{+}}{z-\zeta_{n}}\\
=&M^{+}-I+\frac{i}{z}\sigma_{3}Q_{-}-\mathop{\mathrm{\sum}}\limits_{n=1}^{2N}\frac{\mathop{\mathrm{Res}}\limits_{\overline{\zeta}_{n}}M^{-}}{z-\overline{\zeta}_{n}}
-\mathop{\mathrm{\sum}}\limits_{n=1}^{2N}\frac{\mathop{\mathrm{Res}}\limits_{\zeta_{n}}M^{+}}{z-\zeta_{n}}-M^{+}G
\end{split}\label{Msub}
\end{eqnarray}
The left-hand side of this equation is analytic in $D^{-}$ and is $O\left(\frac{1}{z}\right)$ as $z\rightarrow\infty$, and the sum of the first four terms of the right-hand side is analytic in $D^{+}$ and is $O\left(\frac{1}{z}\right)$ as $z\rightarrow\infty$. The asymptotic behavior of the off-diagonal scattering coefficients implies that $G(x,t,z)$ is $O\left(\frac{1}{z}\right)$ as $z\rightarrow\pm\infty$ and $O(z)$ as $z\rightarrow 0$ along the real axis.

Now we introduce the Cauchy projectors $P_{\pm}$ over $\Sigma$
\begin{equation}
P_{\pm}[f](z)=\frac{1}{2i\pi}\int_{\Sigma}\frac{f(\zeta)}{\zeta-(z\pm i0)}d\zeta,\nonumber
\end{equation}
where $\int_{\Sigma}$ denotes the integral along the oriented contour shown in Fig.1, and the notation $z\pm i0$ indicates that when $z\in\Sigma$, the limit is taken from the left(right) of it. Now recall Plemelj's formulae: if $f^{\pm}$ are analytic in $D^{\pm}$ and are $O\left(\frac{1}{z}\right)$ as $z\rightarrow\infty$, one has $P^{\pm}f^{\pm}=\pm f^{\pm}$ and $P^{+}f^{-}=P^{-}f^{+}=0$. Applying $P^{+}$ and $P^{-}$ to (\ref{Msub}), then we have
\begin{eqnarray}
\begin{split}
M(x,t,z)=&I-\frac{i}{z}\sigma_{3}Q_{-}+\mathop{\mathrm{\sum}}\limits_{n=1}^{2N}\frac{\mathop{\mathrm{Res}}\limits_{\overline{\zeta}_{n}}M^{-}}{z-\overline{\zeta}_{n}}
+\mathop{\mathrm{\sum}}\limits_{n=1}^{2N}\frac{\mathop{\mathrm{Res}}\limits_{\zeta_{n}}M^{+}}{z-\zeta_{n}}\\
&+\frac{1}{2i\pi}\int_{\Sigma}\frac{M^{+}(x,t,\zeta)G(x,t,\zeta)}{\zeta-z}d\zeta, \quad z\in\mathbb{C}\backslash \Sigma.
\end{split}\label{Msolu}
\end{eqnarray}
From (\ref{Msolu}), we deriving the asymptotic behaviors of $M^{\pm}(x,t,z)$ as $z\rightarrow\infty$ as
\begin{eqnarray}
\begin{split}
M(x,t,z)=&I+\frac{i}{z}\Big\{-\sigma_{3}Q_{-}+\mathop{\mathrm{\sum}}\limits_{n=1}^{2N}\mathop{\mathrm{Res}}\limits_{\overline{\zeta}_{n}}M^{-}
+\mathop{\mathrm{\sum}}\limits_{n=1}^{2N}\mathop{\mathrm{Res}}\limits_{\zeta_{n}}M^{+}\\
&-\frac{1}{2i\pi}\int_{\Sigma}M^{+}(x,t,\zeta)G(x,t,\zeta)d\zeta\Big\}+O\big(\frac{1}{z^{2}}\big),
\end{split}\label{Mexp}
\end{eqnarray}
Comparing the (1,2) element of (\ref{Mexp}) with (\ref{muexp}). Then we obtain the formula of the potential $q(x,t)$
\begin{equation}
q(x,t)=e^{2\int_{-\infty}^{x}i\beta(|q|^{2}-q_{0}^{2})dy}\left[q_{-}+i\Big(\mathop{\mathrm{\sum}}\limits_{n=1}^{2N}\mathop{\mathrm{Res}}\limits_{\zeta_{n}}M^{+}\Big)_{1,2}\\
-\frac{1}{2\pi}\int_{\Sigma}(M^{+}(x,t,\zeta)G(x,t,\zeta))_{1,2}d\zeta\right].\label{poten}
\end{equation}

\subsection{N-soliton  solutions}

In this subsection we consider the reflectionless potentials. In this case, the reflection coefficient $\rho(z)=0$,  the potential formula  (\ref{poten}) reduced as
\begin{equation}
q(x,t)=e^{2\int_{-\infty}^{x}i\beta(|q|^{2}-q_{0}^{2})dy}\left[q_{-}+i\mathop{\mathrm{\sum}}\limits_{n=1}^{2N}A_{n}e^{-2i\theta(\zeta_{n})}\mu_{-,11}(\zeta_{n})\right].\label{ppp}
\end{equation}
For convenient, we introduce the quantities
\begin{eqnarray}
a_{j}(x,t,z)=\frac{\tilde{A}_{j}e^{2i\theta(x,t,\overline{\zeta}_{j})}}{z-\overline{\zeta}_{j}}, \quad j=1,\cdots,2N.\nonumber
\end{eqnarray}
Then from (\ref{Msolu}),  we can derive  the results
\begin{subequations}
\begin{align}
&\mu_{-,12}(\overline{\zeta}_{j})=-\frac{iq_{-}}{\overline{\zeta}_{j}}+\mathop{\mathrm{\sum}}\limits_{k=1}^{2N}\frac{A_{k}e^{-2i\theta(\zeta_{k})}}{\overline{\zeta}_{j}-\zeta_{k}}\mu_{-,11}(\zeta_{k})
=-\frac{iq_{-}}{\overline{\zeta}_{j}}-\mathop{\mathrm{\sum}}\limits_{k=1}^{2N}\overline{a}_{k}(\zeta_{j})\mu_{-,11}(\zeta_{k})\\
&\mu_{-,11}(\zeta_{n})=1+\mathop{\mathrm{\sum}}\limits_{j=1}^{2N}\frac{\tilde{A}_{j}e^{2i\theta(\overline{\zeta}_{j})}}{\zeta_{n}-\overline{\zeta}_{j}}\mu_{-,12}(\overline{\zeta}_{j})
=1+\mathop{\mathrm{\sum}}\limits_{j=1}^{2N}a_{j}(\zeta_{n})\mu_{-,12}(\overline{\zeta}_{j}).
\end{align}
\end{subequations}
Substituting the equation into the second one  gives
\begin{equation}
\mu_{-,11}(\zeta_{n})=1-iq_{-}\mathop{\mathrm{\sum}}\limits_{j=1}^{2N}\frac{a_{j}(\zeta_{n})}{\overline{\zeta}_{j}}
-\mathop{\mathrm{\sum}}\limits_{j=1}^{2N}\mathop{\mathrm{\sum}}\limits_{k=1}^{2N}a_{j}(\zeta_{n})\overline{a}_{k}(\zeta_{j})\mu_{-,11}(\zeta_{k}).
\end{equation}

We write this system in matrix form. Let
\begin{equation}
\mathbf{X}=(X_{1},\cdots,X_{2N})^{T}, \quad \mathbf{B}=(B_{1},\cdots,B_{2N})^{T},\nonumber
\end{equation}
where
\begin{equation}
X_{N}=\mu_{-,11}(\zeta_{n}), \quad B_{n}=1-iq_{-}\mathop{\mathrm{\sum}}\limits_{j=1}^{2N}\frac{a_{j}(\zeta_{n})}{\overline{\zeta}_{j}}, \quad n=1,\cdots,2N,\nonumber
\end{equation}
and defining the $2N\times2N$ matrix $A=(A_{n,k})$, where
\begin{equation}
A_{n,k}=\mathop{\mathrm{\sum}}\limits_{j=1}^{2N}a_{j}(\zeta_{n})\overline{a}_{k}(\zeta_{j}), \quad n,k=1,\cdots,2N,
\end{equation}
the system can be rewritten as $M\mathbf{X}=\mathbf{B}$, where $M=I+A=(\mathbf{M}_{1},\cdots,\mathrm{M}_{2N})$.  The system is simply
\begin{equation}
X_{n}=\frac{\det{M_{n}^{ext}}}{\det{M}}, \quad n=1,\cdots,2N,\nonumber
\end{equation}
where
\begin{equation}
M_{n}^{ext}=(\mathbf{M}_{1},\cdots,\mathbf{M}_{n-1},\mathbf{B},\mathbf{M}_{n+1},\cdots,\mathbf{M}_{2N}).
\end{equation}
Finally, upon substituting $X_{1},\cdots,X_{2N}$ into the reconstruction formula (\ref{ppp}),  the $N$-soliton solution of KE equation   can be written compactly as
\begin{equation}
q(x,t)=e^{2\int_{-\infty}^{x}i\beta(|r|^{2}-q_{0}^{2})dy}r(x,t),
\end{equation}
where
\begin{equation}
r(x,t)=q_{-}-i\frac{\det M^{aug}}{\det M},
\end{equation}
and the augmented $(2N+1)\times(2N+1)$ matrix $M^{aug}$ is given by
\begin{eqnarray}
M^{aug}=\left(
\begin{array}{cc}
   0&\mathbf{Y}^{t}\\
   \mathbf{B}&M
\end{array}\right), \quad \mathbf{Y}=(Y_{1},\cdots,Y_{2N})^{T},
\nonumber
\end{eqnarray}
\begin{equation}
Y_{n}=A_{n}e^{-2i\theta(x,t,\zeta_{n})}, \quad n=1,\cdots,2N.
\end{equation}

\subsection{one-Soliton solutions}
In this subsection we mainly consider the one-soliton solution for which the reflection coefficient $\rho=0$.  Take  $q_{0}=1$,
 eigenvalue $z_{1}=i\chi$ ($\chi$=constant value and $\chi>1$)   and $\tilde{A}_{1}=e^{\alpha-i\gamma}$ ($\alpha,\gamma\in\mathbb{R}$). From the $N$-soliton solutions formula, we can obtain the soliton solution
\begin{equation}
q_{1}(x,t)=e^{2\int_{-\infty}^{x}i\beta(|r_{1}|^{2}-q_{0}^{2})dy}r_{1}(x,t),\label{qr}
\end{equation}
where $r_{1}(x,t)$ with the following trigonometric function form
\begin{equation}
r_{1}(x,t)=\frac{\cosh\nu+\frac{1}{2}a_{1}(1+\frac{a_{2}^{2}}{a_{1}^{2}})\sin(\gamma+a_{1}a_{2}t)-ia_{2}\cos(\gamma+a_{1}a_{2}t)}
{\cosh\nu+\frac{2}{a_{1}}\sin(\gamma+a_{1}a_{2}t)},\label{q1}
\end{equation}
and
\begin{eqnarray}
\begin{split}
&\nu=(x+4\beta t)a_{2}+\alpha+a_{0}, \\
&e^{a_{0}}=\frac{a_{1}}{2\chi a_{2}},\\
&e^{-a_{0}}=\frac{2\chi a_{2}}{a_{1}}.\nonumber
\end{split}
\end{eqnarray}
For $\beta=0$, soliton solution (\ref{qr}) convert into the one-soliton solution of focusing nonlinear Schr$\ddot{o}$dinger equation Fig.2 (a). Fig.2 (b) displays the $\beta\neq 0$ case and Fig.2 (c) displays the $\beta\neq 0$ and $\chi\rightarrow 1$ case. The Fig.2 (d) displays the more general situation that is the zero point $z_{1}=i\chi e^{i\alpha}$ where $\chi=2$ and $\alpha=\frac{\pi}{4}$.
\begin{center}
{\includegraphics[scale=0.4]{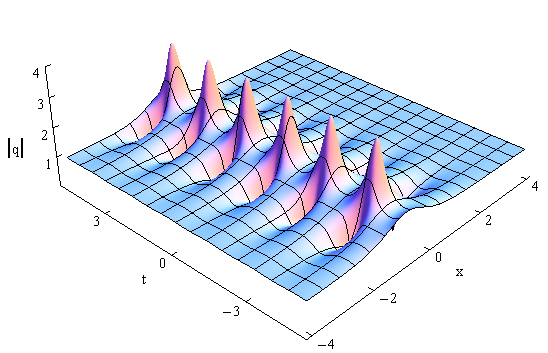}}\hspace{0.3cm}{\includegraphics[scale=0.40]{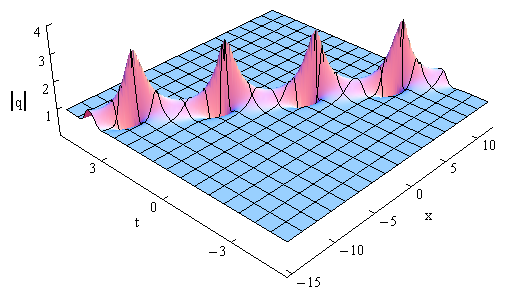}}\\
\vspace{-0.2cm}{\footnotesize\hspace{0.5cm}(a)\hspace{4.5cm}(b)}\\
{\includegraphics[scale=0.4]{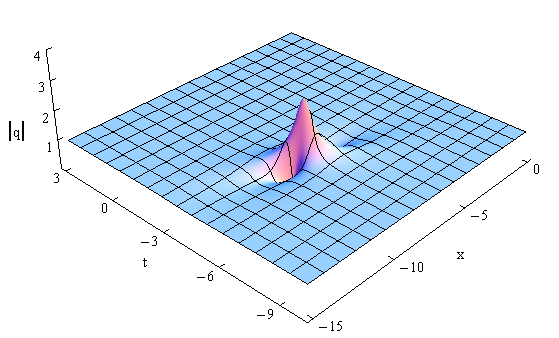}}\hspace{0.3cm}{\includegraphics[scale=0.40]{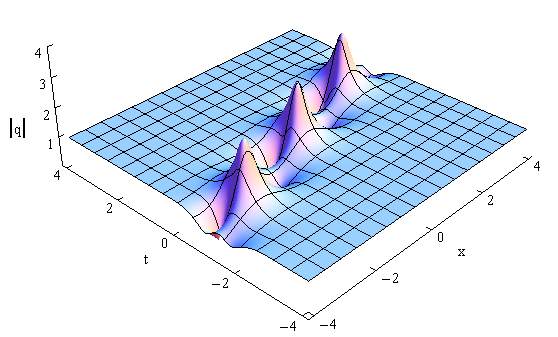}}\\
\vspace{-0.2cm}{\footnotesize\hspace{0.5cm}(c)\hspace{4.5cm}(d)}\\
 \flushleft{\footnotesize {\bf Fig.~2.} (a) The one-soliton solution of equation (\ref{KE}) as $\beta=0$, $\chi=2$.
 (b) The one-soliton solution of equation (\ref{KE}) as $\beta=1$, $\chi=2$. (c) The one-soliton solution of equation (\ref{KE}) as $\beta=\frac{1}{4}$, $\chi=\frac{11}{10}$. (d) The one-soliton solution of equation (\ref{KE}) as $\beta=\frac{1}{4}$, $z_{1}=2ie^{i\frac{\pi}{4}}$.}
\end{center}

\section{Conclusion}
In this work, we investigated the focusing Kundu-Eckhaus equation with nonzero boundary condition at infinity. For $\beta=0$, the soliton solutions are reduced as the soliton solutions of focusing NLS with nonzero boundary conditions. We introduced a transformation, such that  the asymptotic spectral problem with the linear relationship $\mathcal{V}_{\pm}=(2k+2\beta q_{0}^{2})\mathcal{U}_{\pm}$. A appropriate Riemann surface for the single-valued function of the spectral parameter was introduced. Then, the complex $k$-plane transformed into the complex $z$-plane. Unlike the focusing nonlinear Schr$\ddot{o}$dinger equation, the branch cut not on the $\mathrm{Im}k$ axis, but shift $\beta q_{0}^{2}$ along the $x$ axis. The orientation is up to $\beta$.   The nonzero boundary conditions is different from zero boundary conditions mainly reflected in the analytic region and the zero points of $s_{11}$ and $s_{22}$. The analytic region not only involved with the upper-half/lower-half plane, but also involved with the inside/outside of the circle $C_{0}$.

\section*{Acknowledgments}

 This work is supported by the National Science Foundation of China (Grant No.
11671095, 51879045).

\end{document}